\documentclass[runningheads]{llncs}
\usepackage{diagbox}
\usepackage{multirow}
\usepackage{graphicx}
\begin{document}
\title{Physics-informed brain MRI segmentation}
%
%
\author{ 
Pedro Borges\inst{1,2} \and
Carole Sudre\inst{1,2,3} \and
Thomas Varsavsky\inst{1,2} \and
David Thomas\inst{3} \and
Ivana Drobnjak\inst{1} \and
Sebastien Ourselin\inst{2} \and
M Jorge Cardoso\inst{2}}

\authorrunning{P. Borges et al.}
%
\institute{$^1$Department of Medical Physics and Biomedical Engineering, UCL, UK \\ $^2$School of Biomedical Engineering and Imaging Sciences, KCL, UK \\ $^3$Dementia Research Centre, UCL, UK \\
}
\maketitle              
\begin{abstract}
Magnetic Resonance Imaging (MRI) is one of the most flexible and powerful medical imaging modalities. This flexibility does however come at a cost; MRI images acquired at different sites and with different parameters exhibit significant differences in contrast and tissue appearance, resulting in downstream issues when quantifying brain anatomy or the presence of pathology. In this work, we propose to combine multiparametric MRI-based static-equation sequence simulations with segmentation convolutional neural networks (CNN), to make these networks robust to variations in acquisition parameters. Results demonstrate that, when given both the image and their associated physics acquisition parameters, CNNs can produce segmentations that exhibit robustness to acquisition variations. We also show that the proposed physics-informed methods can be used to bridge multi-centre and longitudinal imaging studies where imaging acquisition varies across a site or in time. 

\end{abstract}
\section{Introduction}
Magnetic Resonance Imaging (MRI) is a widespread, non-invasive, non-ionizing medical imaging technique. It is capable of imaging any part of the body to produce three dimensional anatomical and functional reconstructions, excelling at soft tissue contrast. MRI is therefore aptly suited for looking at pathological changes in the brain, such as tissue atrophy or lesions. However, large scale studies that rely on multiple scanners from different manufacturers suffer from site and hardware-dependent variabilities in the acquired data \cite{ScannerVariability}. Without means to account for these differences, this variability impacts our ability to conduct multi-centre analyses and extract meaningful and reproducible biomarkers \cite{Frisoni2017}. Further challenges are encountered in longitudinal studies since scanner and sequence protocol changes cause inconsistencies in patient imaging that make disease evolution impossible to quantify. 

One issue caused by the non-quantitative nature of MRI images is that algorithms are often unable to deal with different sequence parameters. For example, if MR images are acquired with small differences in acquisition, segmentation algorithms often produce disparate results, exhibiting apparent growth/shrinkage of regions of interest \cite{ProtocolVariability}: this is caused by signal intensity changes that depend on the tissue. Dementia is an example of a condition in which MR imaging biomarkers such as cortical atrophy or hippocampal volume can be used for the diagnosis of the condition. Due to the difficulty in disentangling imaging physics and underlying anatomy, even trained clinicians may fail to account for protocol differences. This is explained by the fact that the information available to the users is limited to voxel intensities, which are unreliable, and \textit{a priori} knowledge of brain morphology, which is not subject-specific.

Current methods to mitigate these effects are cumbersome and imperfect. They normally rely on either a per-site analysis followed by a joint meta-analysis \cite{vanErp2015} or the use of statistical models with linear covariates \cite{Ashburner2000}, which provide computationally efficient yet often inaccurate means of standardization. When attempted, correction for scanning variability can either be made at the tissue-class or the voxel level. The former may be less robust due to the lack of granularity whilst the latter is highly susceptible to errors induced by the required registration to a group-wise space.

Gaining robustness to pulse sequences has been investigated in \cite{jog2018pulse} implicitly, where parameter estimation combined with simulation as an augmentation to a segmentation task is employed. Note however that parameters are bulk-assigned to segmentation maps and that the segmentation network is not physics aware.

This work aims to address the issue of acquisition-induced biomarker extraction variability (resulting from imprecise segmentations) through the use of a novel physics-informed convolutional neural network, where segmentation is used as a pretext task. Parametric tissue maps, together with sequence simulating models, are used to generate realistic samples as if they were acquired with different physics parameters. The importance of MRI sequence simulators is crucial here --- they can be used to train machine learning algorithms to learn the features of scanners or sequence parameters without having to expend a vast amount of resources to acquire
real data, while realistically reproducing an MRI scan. For computational reasons, we propose to use a static equation-based simulation approach, which makes use of simplified imaging equations to most efficiently generate sufficiently large and varied datasets required for this undertaking. By providing both simulated samples and associated physics parameters as training data for a neural network, we can demonstrate that these networks can produce segmentations that are robust to MRI physics variations because they are explicitly learning how the physics interacts with image contrast. 

\section{Methodology}
Here we describe how physics-based image synthesis is combined with the proposed physics-informed architecture to achieve more consistent segmentations.

\subsection{MRI simulation methods}
The signal intensity obtained in MR images results from the non-linear interaction between tissue properties and parameters associated with a specific acquisition sequence. In this work, we use a simplified, yet robust, simulation model by which static singular equations are employed for each simulated sequence previously employed by \cite{JOG}. These equations take as input the tissue properties and MR sequence parameters and produce an image of the corresponding sequence. The assumption that signal intensity at any given voxel is based on tissue, sequence, and scanner properties is made here. Contrarily to more complex MR simulators, the proposed model assumes the signal does not change in time, i.e. the model described as `static'. In this work, we focus on two widely used gradient echo T1-weighted sequences, namely 3D spoiled Gradient echo (SPGR) and 3D magnetization prepared gradient echo (MPRAGE). For both sequences, $PD(x)$, $T_1(x)$ and $T_2^{*}(x)$ are respectively the proton density, $T_1$ value and $T_2^{*}$ value of the tissue at position $x$,  $\theta$ is the sequence flip angle, $TR$  the relaxation time and $TE$ the echo time.
For the SPGR sequence, the static equation derived by Jog \textit{et al.} \cite{JOG} expresses the voxel intensity $b_S(x)$ at position $x$ as
\begin{equation} \label{eq:eqErnst}
b_S(x) = G_SPD(x)sin{\theta}\frac{1-e^{-\frac{TR}{T_1(x)}}}{1-\cos{\theta}e^{-\frac{TR}{T_1(x)}}}e^{-\frac{TE}{T_2^{*}(x)}},
\end{equation}
where $G_S$ is the scanner gain.
Similarly, the static equation for MPRAGE sequences describes the intensity $b_M(x)$ at position $x$ as

\begin{equation} \label{eq:eqMPRAGE}
b_M(x) = G_MPD(x)\Bigg(1-\frac{2e^{\frac{-TI}{T_1(x)}}}{1+e^{\frac{-(TI+TD+\tau)}{T_1(x)}}}\Bigg),
\end{equation}
\newline
where $G_M$ is the scanner gain, $TD$ the delay time, and $\tau$ the slice imaging time.

\begin{figure}
\centering
\includegraphics[width=0.310\textwidth]{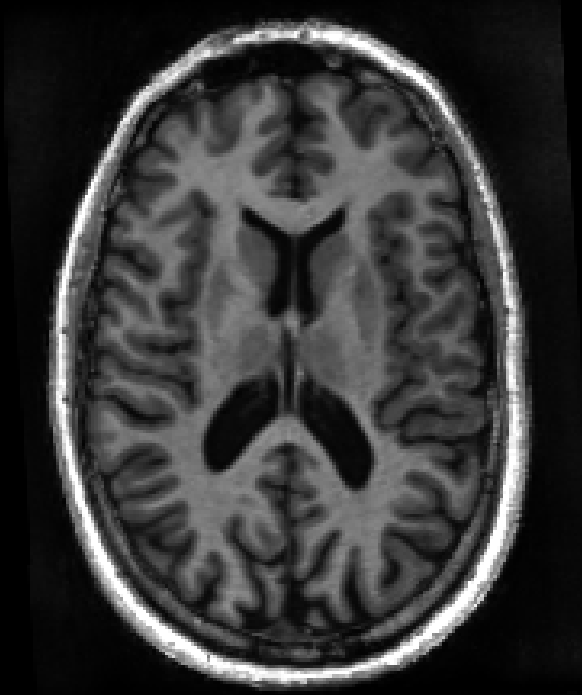}
\includegraphics[width=0.316\textwidth]{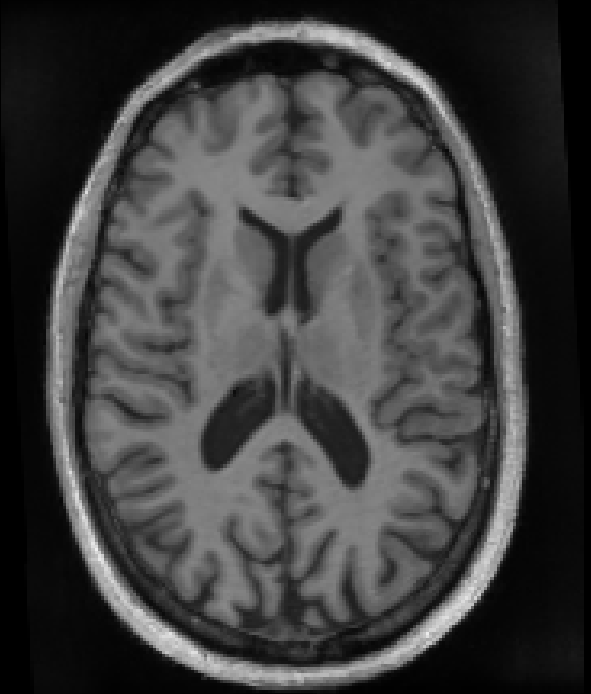}
\caption{Left: Middle axial slice of simulated MPRAGE volume with TI=600ms. Right: Middle axial slice of simulated MPRAGE volume with TI=1200ms. Note the difference in tissue contrasts, particularly between white and cortical grey matter.} \label{fig:figContrasts}
\end{figure}

A variable inversion time (TI) results in images of differing contrasts owing to the non-linear signal scaling of the different tissue types. Figure ~\ref{fig:figContrasts} illustrates the phenomenon, showcasing two axial slices of MPRAGE simulations with TIs of 800 and 1200ms, respectively, simulated from the same set of parametric maps.

\subsection{Physics-aware CNNs for image segmentation}
We propose to inject the sequence parameters of an input image in a CNN network by the addition of a fully connected layer. Since a strong correlation between segmentation volumes and parameter choice can be observed, we expect the introduction of the physics parameters to allow such a network to account for the physics induced appearance variability and attain a more consistent, unmarred segmentation. 

\subsection{Network architecture}
\begin{figure}[!b]
\centering
\includegraphics[width=0.9\textwidth]{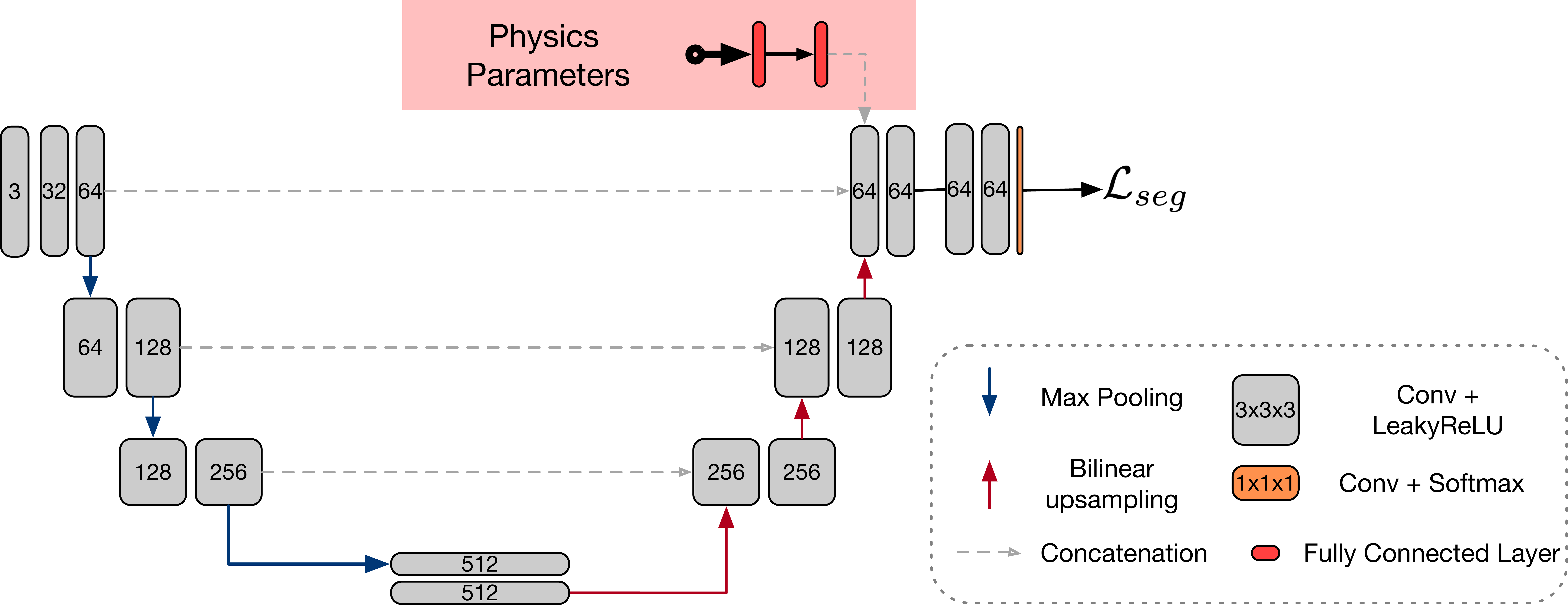}
\caption{Diagramatic representation of the proposed network. The novel contribution is primarily the pink box (physics-aware subnet) and the simulation framework used to train the model.} \label{fig:HRN}
\end{figure}

We used the 3D U-Net architecture as described in \cite{3dunet} as a starting point for the proposed physics injected network. Our proposed architecture (Figure \ref{fig:HRN}) adds an adjacent branch (physics branch), boasting two fully connected layers of ten neurons each which have as input an N-dimensional vector. This vector consists of the variable physics parameters used to generate the image to be segmented, as well as negative exponentiation of said parameters. The latter is included as a means of making the network privy to the underlying simulation process. This branch is connected to the network via a concatenation operation that broadcasts the branch's output as additional channels immediately following the final shortcut connection of the base 3D U-Net architecture.\looseness=-1

\section{Experiments}
\subsection{Creation of physics-based gold standard segmentation}
It is important to reiterate the primary goal of this work, which is to mitigate the detrimental effects of imaging parameter choices on the consistency of MRI image segmentations. As such, we are primarily concerned with achieving segmentations that are consistent across acquisitions, more so than approaching a hypothetical `ground truth' segmentation. With this in mind, and as proper ground truth tissue segmentations are `illusory' in brain imaging (i.e. humans often disagree, and image resolution is insufficient), we propose to create a gold standard reference segmentation that is consistent and stable across subjects, i.e. one that is precise, concordant and systematic, but not necessarily accurate. 

To achieve this, and owing to the T1-W nature of the simulations we propose a "Physics Gold Standard", whereby we make use of the quantitative $R_1$ maps for this purpose. By assuming that tissues can be parameterised by normal distributions, with mean and standard deviations equivalent to literature $R_1$ values, we can generate segmentation maps directly from quantitative acquisitions, thereby being largely independent of acquisition physics. Because there can be some significant variation in quantitative maps, we opt to use parameters derived from works whose multiparametric map creation protocol most resembles our own, with an increased standard deviation to account for some additional variability. To this end, the $R_1$ values we choose are 0.683 $\pm$ 0.080 ms for grey matter, 1.036 $\pm$ 0.080 ms for white matter, taken from \cite{PGS_Params}. We note that \cite{PGS_Params} do not quote values for CSF, but as per \cite{CSF_Params} which cites multiple other studies on this matter, the $R_1$ values for CSF are largely independent of acquisition parameters. Due to this, and the fact that we only focus on grey and white matter segmentation consistency in this work, we opt to model our normal CSF distribution with 0.240 $\pm$ 0.03 ms, one of the cited values that most closely matched the $R_1$ CSF measurements in our maps. Using this "Physics Gold Standard" we can model each tissue in a probabilistic, and more anatomically grounded, manner without having to concern ourselves with the inherent bias that would be associated with choosing a more typical "ground truth".

\subsection{Datasets}
27 multiparametric volumes from an early onset Alzheimer's disease dataset containing both patients and controls were used for simulation. The maps consist of $R_1$ (longitudinal magnetisation relaxation rate), $R_2^{*}$ (effective transverse magnetisation relaxation rate), proton density (PD), and magnetisation transfer (MT). We make use of the former three for the simulations. These maps are acquired via three 3D multi-echo FLASH (fast low angle shot) acquisitions further described in \cite{helms2009increased}. All subjects were rigidly registered to MNI space before their use in simulations. 

As a real-world data example, we used a subset of the SABRE dataset consisting of data from 22 subjects drawn from an elderly population with high cardiovascular risk factors, where each subject was imaged within the same scanning session using two different T1-W MPRAGE protocols and one Turbo Spin-Echo protocol. We use only the paired MPRAGE images in this work. Mid-space (so as not to bias the registration towards either acquisition protocol) intra-subject registrations are carried out to resample these images to be 1mm isotropic.

\begin{table}[t!]
\caption{Mean dice scores of GIF, base 3D U-Net and 3D U-Net-Physics methods on segmentation task across inference subjects. All dice scores are estimated against a Physics Gold Standard.}\label{tab:tabDice}
\centering
\begin{tabular}{| c | c | c | c | c |}
\hline
\multirow{3}{*}{Experiments} & \multicolumn{4}{|c|}{Sequences} \\ \cline{2-5}
& \multicolumn{2}{|c|}{MPRAGE} & \multicolumn{2}{|c|}{SPGR} \\ \cline{2-5}
& GM & WM & GM & WM \\
\hline
GIF & 0.851 $\pm$ 0.020 & 0.919 $\pm$ 0.009 & 0.916 $\pm$ 0.010 & 0.847 $\pm$ 0.018\\
Base & 0.904 $\pm$ 0.024 & 0.943 $\pm$ 0.012 & 0.946 $\pm$ 0.017 & 0.902 $\pm$ 0.021\\
Physics & 0.910 $\pm$ 0.017 & 0.948 $\pm$ 0.009 & 0.948 $\pm$ 0.019 & 0.907 $\pm$ 0.015 \\
\hline
\end{tabular}
\end{table}

\subsection{Simulation experiment and results}
For each of the 27 subjects acquired with a multiparametric acquisition, we simulated 121 MPRAGE volumes with TIs between 600 and 1200 ms (5 ms increments) with constant TD of 600 ms and constant $\tau$ of 10 ms. These values were chosen by extending the optimised range of 900 - 1200 ms found in \cite{MPRAGEParams}. Similarly, 121 SPGR volumes were simulated per subject sampling randomly from the parameter space spanning TR between 15 and 100 ms, TE between 4 and 10 ms, and FA between 15 and 75 degrees. These values were chosen according to the typical values explored in \cite{JOG} for SPGR sequences. For each subject, a single "Physics Gold Standard" segmentation was used across the associated synthesized images.

\subsection{Robustness to acquisition parameters}
We train our network on 3D 96x96x96 randomly sampled from the simulated volumes. For the MPRAGE volumes, TI and $e^{-TI}$ are passed as a vector into the physics branch while for SPGR volumes a vector containing TR, TE, FA, $e^{-TR}$, $e^{-TE}$, and $sin(FA)$ is passed as input to the physics branch. Subjects were randomly split between training (2420 simulated volumes over 19 subjects), validation (242 simulated volumes over two subjects) and testing (726 simulated volumes over six subjects). Networks were trained until convergence, as defined per performance on the validation set when over 1000 iterations have elapsed without decreases in the loss (a probabilistic version of the dice loss \cite{milletari2016v}), using NiftyNet \cite{Gibson2018}, a deep learning framework designed for medical imaging.

\begin{figure}[b!]
\centering
\includegraphics[width=1.0\textwidth]{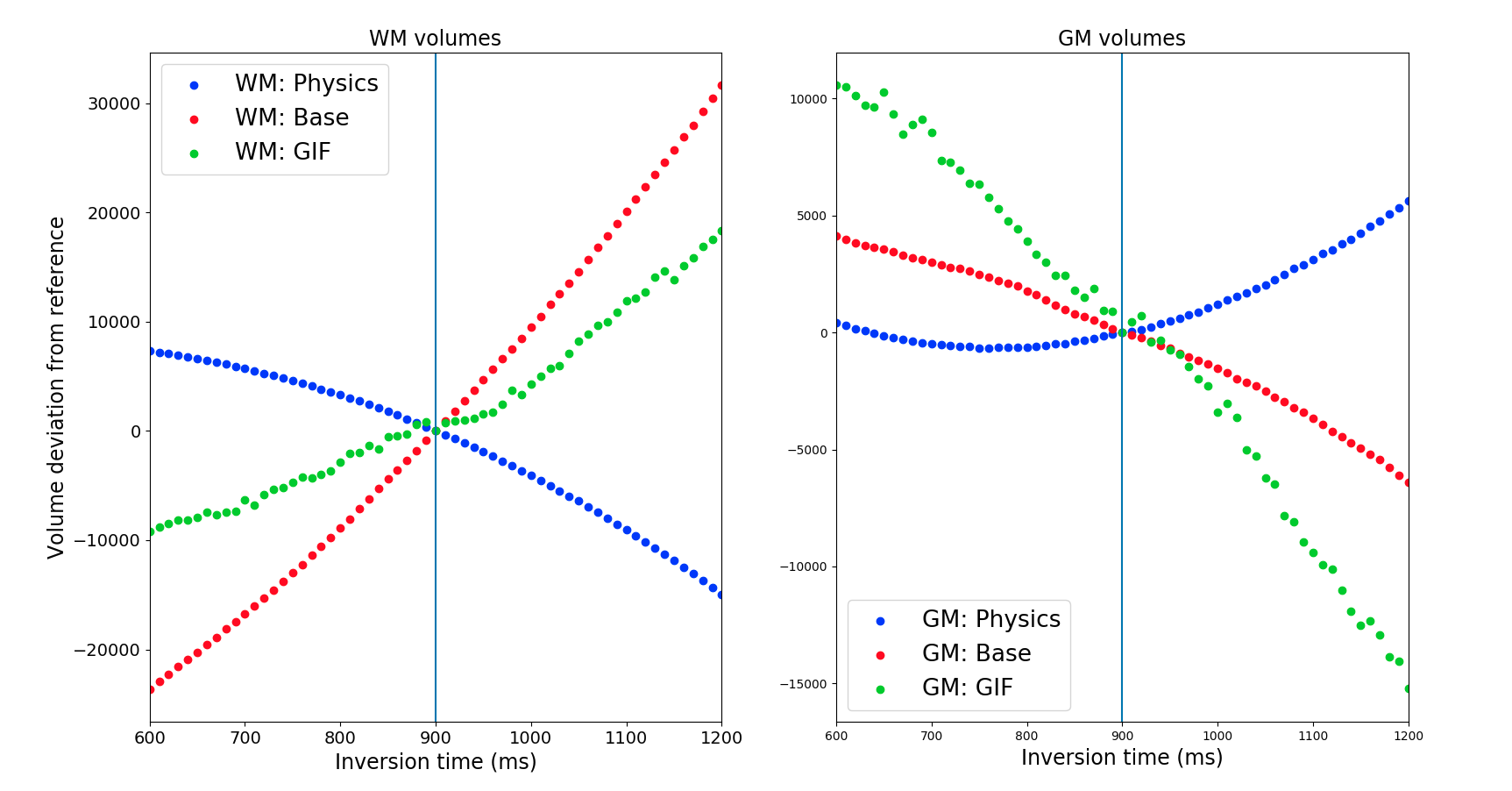}
\caption{Comparing the (left) WM and (right) GM volume consistency for simulated data within the 600 - 1200 TI range over three methods (GIF, Baseline, Physics). Volumes are presented as a deviation from the volume at TI=900ms.} \label{fig:figConsistency}
\end{figure}

In a first instance, we compare the performance of 3D U-Net-Physics with that of base 3D U-Net (trained simply by excluding the physics branch), as well as GIF \cite{Cardoso2015}, a segmentation software based on geodesical information flows. Segmentation stability for the two tissue classes is assessed via measures of the coefficient of variability within the set of synthesized data for each test subject. Results, presented in Table \ref{tab:tabCoV} show that for MPRAGE simulation, the model enhanced with the Physics branch provides more stable volume estimates for the two tissue classes. Table \ref{tab:tabDice} shows the dice scores for the segmentations compared to the "Physics Gold Standard". It is apparent that "accuracy"-wise all methods perform similarly. A sign-rank test is carried out to calculate the p-value of the Physics and Base method dices, resulting in p \textless 0.0001, indicating the Physics model's higher performance is statistically significant. The larger gap in performance between the CNN and GIF methods can likely be attributed to differences exhibited between our "Physics Gold Standard" and more typical segmentation ground truths, the latter being closer to what GIF might output. 
For a single subject, we plot in Figure \ref{fig:figConsistency} the resulting variations in extracted volumes for a range of TI compared to the standard obtained at 900 ms for WM (left) and GM (right). The introduction of the physics branch to the model appears to noticeably reduce the variability observed when using a non-physics aware network. 


\begin{table}[t!]
\caption{Mean coefficients of variability resulting from GIF, 3D U-Net and 3D U-Net-Physics methods on segmentation task across inference subjects}\label{tab:tabCoV}
\centering
\begin{tabular}{| c | c | c | c | c |}
\hline
\multirow{3}{*}{Experiments} & \multicolumn{4}{|c|}{Sequences} \\ \cline{2-5}
& \multicolumn{2}{|c|}{MPRAGE} & \multicolumn{2}{|c|}{SPGR} \\ \cline{2-5}
& GM & WM & GM & WM \\
\hline
GIF & 0.00619 & 0.00783 & \textbf{0.00114} & 0.00091 \\
Base & 0.00332 & 0.01491 & 0.00531 & 0.00160 \\
Physics & \textbf{0.00244} & \textbf{0.00405} & 0.00425 & \textbf{0.00083} \\
\hline
\end{tabular}
\end{table}

\subsection{Application to a data bridging study}

\begin{figure}[t!]
\centering
\includegraphics[width=0.65\textwidth]{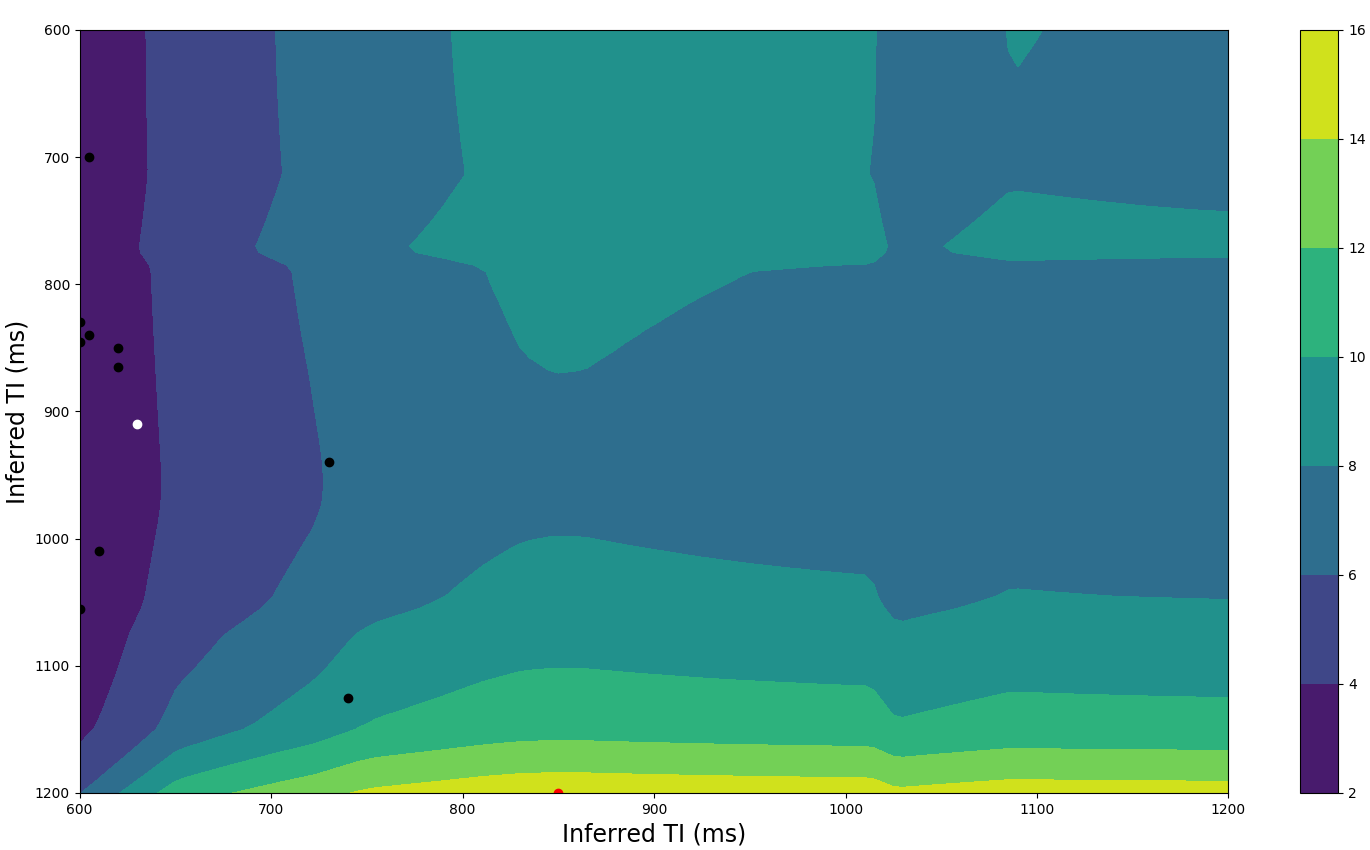}
\includegraphics[width=0.65\textwidth]{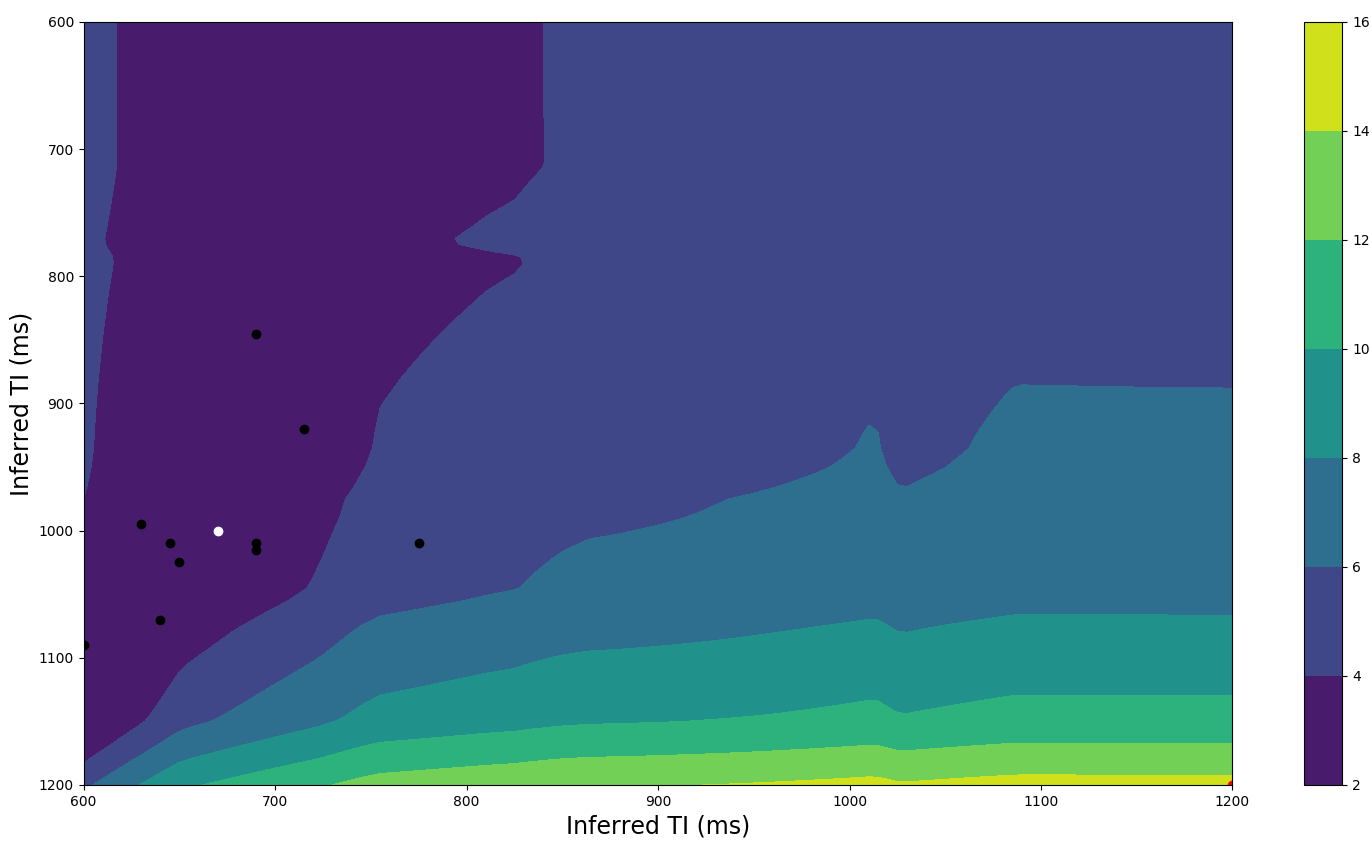}
\caption{Top: GM contour of mean percent volume differences across a "training" (12 subjects) subset of SABRE paired dataset.
Bottom: WM contour of mean percent volume differences across a "training" (12 subjects) subset of SABRE paired dataset.
Black points: Ideal params for each of 10 "test" subjects. White point: Point of volume pair minimisation for "training" subjects.} \label{fig:figContours}
\end{figure}

MRI acquisitions are often updated due to hardware or sequence changes, harming our ability to perform ongoing clinical research studies and to compare patients longitudinally. This experiment aims to test if the proposed method could be used to compensate for acquisition parameter updates from an image segmentation biomarker. To do this, we use the 3D-UNet-Physics network trained in the previous section and applied it to the Bridge SABRE dataset, where each subject has a pre and a post parameter update MRI acquisition. Despite images being acquired back-to-back, these images exhibit significant differences owing to differing scanning protocols. As MRI acquisition parameters were not available for this data, we experimentally found an optimal TI parameter that, if applied to the pre-update and post-update acquisition parameters, would make the segmentations between those as close as possible. To achieve this, we partitioned the data into two sets, a "training" set with 12 image pairs, and a "test" set with 10 image pairs. We segment all training set pairs with different hypothetical TI values (i.e. run inference on our physics informed network for each image 121 times, passing a different value of TI from the 600 - 1200 ms TI range at each pass), and plot the similarity (volume-wise) between TI pairs in Fig. \ref{fig:figContours} as a contour. This plot shows us how the similarity between GM and WM between the pairs changes according to the TI passed at inference to segment the images. The white point in the figure represents the pre/post TI pair that minimizes the volume difference between training set examples. Finally, we apply this TI parameter pair to the hold-out test set and test if the output segmentations were bridged. Similarly to the white point, the black points indicate the TI pairing that provided the smallest difference in tissue segmentation volumes for each of the 10 subjects. Qualitatively, we observe clustering of these black points around the white point, indicating that this choice of parameters generalizes to the holdout standardization subjects, lending credence to the model's ability to account for the imaging process. 

To quantitatively compare to a baseline method, we ran GIF on both pre and post upgrade scans, estimating the WM and GM volumes for each subject - here denoted GIF-Uncorrected. We then find a linear mapping between pre- and post-sequence upgrade volumes on the training set as a form of correction/bridging, and apply it to the hold-out test set - here denoted GIF-Corrected. GIF-Uncorrected exhibits a volumetric difference over tissues of 5\%. We found that the linear correction method applied to GIF resulted in a mean volumetric segmentation difference of 3\% over tissues. When using the physics model (i.e. choosing the aforementioned ideal parameter pairs denoted by the white point in \ref{fig:figContours}), this value was found to be 2\%. While the differences do not seem large, any improvements are beneficial for the sample size for a hypothetical trial. Consistency could have been potentially improved further if true physics parameters were available, rather than experimentally found. 

\section{Discussion and Conclusions}
This work aimed to address the problem of imaging induced variability in biomarker extraction by constructing networks that are privy to those imaging parameter choices. We show that, for MPRAGE-type sequences, our method outperforms its alternatives, particularly with regards to grey matter. For SPGR we note that performance is more tissue dependent, with small, but significant, decreases in grey matter segmentation consistency compared to competing methods but presenting improvements in white matter consistency. Despite this, it is worth noting that our physics informed network always outperforms its base counterpart, lending credence to the notion that these external parameters can always be leveraged towards some gain in segmentation consistency. 

Further work is needed to ascertain the ideal architectural setup for passing external parameters. While the proposed construction performs well for MPRAGE sequences, the results suggest that this might not be the case for others, especially if multiple parameters are involved: The network may require a greater capacity to leverage the additional information.

Results on real data show that the proposed method improved the consistency of estimated volumes, even when the acquisition parameters were found experimentally. Future work will test the proposed method on a more diverse range of protocols and will apply it to larger bridge studies where significant improvements stand to be made.

\subsubsection{Acknowledgements}
We gratefully acknowledge the support of NVIDIA Corporation with the donation of one Titan V. This project has received funding from Wellcome Flagship Programme (WT213038/Z/18/Z) and Wellcome EPSRC CME (WT203148/Z/16/Z).

\bibliographystyle{splncs04}
\bibliography{example}
\end{document}